\documentclass[twocolumn,notitlepage,floats,nofootinbib,amsmath,amssymb,aps,prd]{revtex4-1}

\usepackage{amsmath,amssymb,amsfonts,latexsym,cancel,amsthm}

\usepackage{mathtools}
\usepackage{graphicx}
\usepackage{color}
\usepackage{amsbsy}
\usepackage{hyperref}
\usepackage{tikz}
\usepackage{seqsplit}
\usepackage{comment}
\usepackage{enumerate}

\newcommand{\be}{\begin{equation}}
\newcommand{\beq}{\begin{equation}}
\newcommand{\ee}{\end{equation}}

\def\bea {\begin{eqnarray}}
\def\eea {\end{eqnarray}}

\def\dd{{\rm d}}

\begin{document}

\title{Quantum gravitational stellar evolution beyond shell-crossing singularities}

\author{Micha{\l} Bobula} \email{michal.bobula@uwr.edu.pl}
\affiliation{University of Wroc{\l}aw, Faculty of Physics and Astronomy, Institute of Theoretical Physics,
Plac Maksa Borna 9, 50-204 Wrocław, Poland}

\author{Francesco Fazzini} \email{francesco.fazzini@unb.ca}
\affiliation{ Friedrich-Alexander-Universität Erlangen-Nürnberg, \\
Institute for Quantum Gravity, 
Staudtstr. 7, 91058 Erlangen, Germany}
\begin{abstract}
\noindent
Models of effective stellar collapse inspired by loop quantum gravity predict a bounce when the stellar energy density reaches the Planck scale, typically followed by the formation of shell-crossing singularities. This work aims to extend the spacetime beyond these singularities by employing a Hamiltonian formulation of the Darmois-Israel junction conditions, treating the singularity as a non-isolated thin dust shell. By construction, the shell's motion remains timelike throughout the entire evolution, regardless of the amount of initial stellar mass, and the induced metric on the shell remains continuous. The resulting stellar evolution produces an inter-universal wormhole, analogous to the simpler Oppenheimer-Snyder scenario. The proposed approach provides a general framework for any effective (or classical) theory of stellar collapse characterized by shell-crossing singularities.

\end{abstract}

\maketitle

\section{Introduction}
\noindent In standard general relativity, spherically symmetric stellar collapse predicts the formation of a central singularity, provided that the matter pressure is insufficient to halt the collapse before the star reaches its Schwarzschild radius, in accordance with Penrose’s singularity theorem \cite{Penrose:1964wq}. The emergence of a physical, Tipler-strong \cite{Tipler1977,Krolak1986}, spacelike singularity (in the chargeless case) precludes the extension of the spacetime beyond the singular hypersurface, resulting in geodesic incompleteness and the breakdown of the classical theory. 

Despite being a fundamental limitation to the description of spacetime, central singularities are not the only type of singularities that can arise during gravitational collapse. Indeed, it has been demonstrated that in both the dust case \cite{Hellaby:1985zz,Szekeres:1995gy,Nolan_2003} and the case with pressure \cite{MullZummHagen1973}, another class of singularity may emerge, commonly referred to as a \emph{shell-crossing} singularity (SCS) \cite{Szekeres:1995gy}. However, unlike the unavoidable central singularities, it is easy in the classical theory to identify initial density profiles that do not develop such singularities \cite{Hellaby:1985zz}; this is one of the primary reasons they are not of central interest in classical contexts.

The physical interpretation of these singularities is well-understood: one may conceptualize the collapsing star as a set of concentric matter shells, evolving either independently (as in the dust case) or under the influence of adjacent layers (as in the case with pressure). If two matter shells intersect during their collapse, a physical singularity forms, as the volumetric energy density diverges at the spacetime point of the crossing. This type of singularity is less prominent and, simultaneously, less problematic than the central one. Specifically, although radial tidal forces diverge in the proximity of shell-crossing singularities, the tidal radial geodesic deviation remains finite \cite{Szekeres:1995gy}, suggesting the possibility of extending the spacetime beyond the singular point. Consequently, physical volumes enclosing the shells do not vanish or diverge, in contrast to the behavior at the central singularity. For these reasons, shell-crossing singularities are categorized as Tipler-weak \cite{Nolan2001}.

The possibility of extending spacetime beyond these weak singularities has been extensively investigated in the literature. The most straightforward and physically motivated approach to such an extension arises from recognizing that the equation of motion for dust collapse---the Lemaître-Tolman-Bondi (LTB) equation---in Painlevé-Gullstrand (PG) coordinates take the form of a non-linear hyperbolic conservation law. At shell-crossing singularities, the corresponding partial differential equations (PDEs) exhibit characteristic crossings \cite{Nolan_2003}. Consequently, within the mathematical framework of PDEs, it is possible to study the evolution of the system beyond these crossings through the use of weak solutions \cite{leveque-92}. Under this framework, the multi-valued behavior of the gravitational field 
(specifically the extrinsic curvature) at the characteristic crossing is interpreted as a spatial discontinuity—a physical shock\footnote{{In this work, the term shock refers to two qualitatively distinct scenarios of discontinuous gravitational field: (i) where both the induced metric and the extrinsic curvature are discontinuous across a surface, and (ii) where the induced metric is continuous while the extrinsic curvature is discontinuous. In both cases, these discontinuities are treated in a distributional sense, though we shall focus exclusively on the latter.} }. The evolution of this shock, as described by the weak solution, is governed by the integral form of the original PDE. 

While this approach has successfully captured shock dynamics in various fields, ranging from fluid dynamics \cite{Chavanis:2019auu} to condensed matter \cite{Hoefer_2006}, its application to general relativity must address the inherent subtleties of Einstein's theory. Specifically, a weak solution is a solution to an integral equation in which the time variable is implicitly assumed to be continuous across the shock. Although this is not a limitation in fluid dynamics or condensed matter—where time flows uniformly in a Newtonian sense—it represents a significant constraint within general relativity \cite{Fazzini2025}. Indeed, when a gravitational shock forms due to the collision of adjacent matter shells, the spacetime geometries on either side of the shock are generally distinct, even when a weak solution with a continuous induced metric is selected \cite{Husain2025}. This implies that the flow of time inside and outside the shock differs, rendering the requirement of a continuous time variable physically unjustifiable. Furthermore, this assumption precludes the possibility of imposing a timelike character on the shock hypersurface \cite{Fazzini2025} (and thus on the thin-shell hypersurface formed at the shock), which is necessary for a physically realistic description of the matter particles composing the shell. Consequently, the weak solution lacks control over the signature of the shock hypersurface, potentially leading to shock dynamics that change character—from timelike to null or spacelike—during the evolution \cite{Husain2025}.

The most physically sound method for treating the shocks emerging from shell-crossing singularities is through the Darmois-Israel junction conditions. These conditions ensure the timelike behavior of the shock at all times and allow for the Einstein field equations to be satisfied in a distributional sense \cite{Israel1966,Poisson2004Toolkit}. Such extensions for shell-crossing singularities have previously been proposed in classical cosmological contexts, for instance, to investigate the evolution of cosmological voids \cite{Maeda1983,Sakai1993} and structure formation \cite{Sakai:1999xx}.

A primary motivation for developing a theory of quantum gravity is the construction of a stellar collapse model that avoids central singularities, leading instead to regular black holes and non-singular collapse. Among various approaches, loop quantum gravity (LQG) remains a leading candidate \cite{Rovelli:2004tv, Thiemann:2007pyv}. Although deriving stellar collapse dynamics from the full theory remains a formidable task, effective models inspired by its fundamental features have been developed in recent years (see e.g., \cite{Kelly:2020lec, AlonsoBardaji2024, Bojowald2025}). While their precise relationship with the full theory is still being clarified, most of these models resolve the central singularity for arbitrary initial conditions while preserving the correct classical limit. Within this class, only one model to date reduces to standard effective loop quantum cosmology (LQC) in its homogeneous sector \cite{Kelly:2020lec, Husain:2021ojz, Husain2022, Giesel:2023hys,Giesel:2025kdl}. This model prevents the formation of a central singularity by generating gravitational repulsion when the stellar energy density reaches the Planck scale, resulting in a bounce of the constituent matter shells. However, unlike in ordinary LQC---where a perfect homogeneity is usually imposed, the inhomogeneities in the density profiles cause shell-crossing singularities to emerge almost immediately (on a Planckian timescale) following the bounce. In contrast to the classical case, the formation of these singularities post-bounce appears unavoidable for any physically reasonable initial conditions, both for dust \cite{Fazzini:2023ova, Cipriani:2024nhx} and fluid with pressure \cite{Cafaro:2024lre}. { Moreover, the analysis presented in \cite{Bobula:2024chr} examined the causal structure of the characteristic curves (LTB shells). It was shown that, in the causal future beyond the first shell-crossing singularity, the shells overlap, thereby encoding different spacetime geometries at the same spacetime points.} Consequently, extending the spacetime beyond these singularities is of paramount importance for this effective model.

The initial attempt to extend the effective spacetime beyond these singularities was conducted in \cite{Husain2022}, utilizing weak solutions with a discontinuous induced metric at the shock. However, the resulting shock dynamics suffer from the same issues encountered in the classical case. Specifically, the discontinuity of the induced metric precludes the imposition of the first Israel junction condition. This lack of metric continuity on the shell prevents a unique definition of geometric quantities, such as the shock hypersurface signature, leading to inconsistent signatures when computed using the interior versus the exterior metric. The metric discontinuity {of this type} was subsequently mitigated in \cite{Liu:2025fil}, where weak solutions with a continuous metric at the shock were constructed. Nevertheless, the enforced continuity of the Painlevé-Gullstrand (PG) time still prevents the shock from exhibiting subluminal behavior, resulting in superluminal motion as in the classical approach. Another proposed effective solution is presented in \cite{Sahlmann2025}, where the PG time is permitted to be discontinuous; however, the induced metric remains discontinuous, and the dynamics remain superluminal. The latter demonstrates that allowing PG coordinates to be discontinuous at the shock is a necessary, but not sufficient, condition for well-defined dynamics. The sufficient condition is the requirement that the normalized four-velocity of the shock, as measured by both the interior and exterior metrics, is timelike. This requirement automatically enforces the first Darmois-Israel junction condition \cite{Fazzini2025}.

\begin{figure*}
    \centering
    \includegraphics[width=0.9\linewidth]{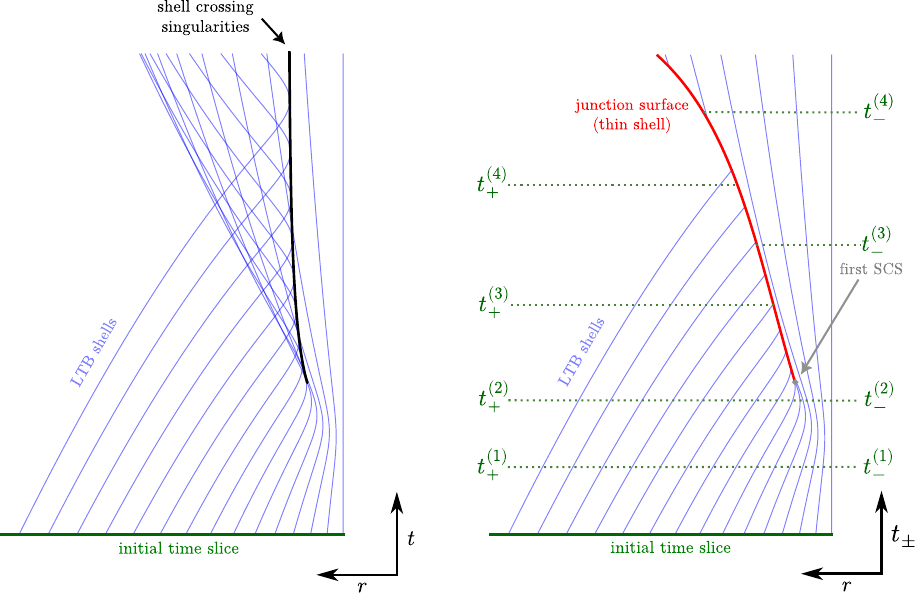}
    \caption{ Schematic plots illustrating two {dynamical extensions of LTB solutions beyond shell-crossing singularities} (left and right diagrams) of the same solutions to the LTB equations of motion are shown, where the trajectories of the LTB shells are evolved from a initial time slice and displayed on the areal radius $r$–proper time $t$ plane.\\
\emph{Left diagram.} The dynamical extension in which shell crossings among the LTB shells are allowed and therefore manifest. In this picture, different shells may intersect in the $(r,t)$ plane, leading to the formation of shell-crossing singularities. In particular, we can distinguish between a family of shells that undergo an early bounce and another family that bounces at later times. An inconsistency arises because the latter family encodes a spacetime geometry (metric) that differs from that of the former family in the region where their domains of validity overlap. In other words, at spacetime points $(r,t)$ within the overlapping region, two distinct spacetime metrics are assigned by the LTB shells belonging to the respective families. Importantly, the proper time $t$ on the LTB shells remains continuous across the shock surface (shown in black). The picture of this type was explicitly obtained in \cite{Fazzini:2023ova, Bobula:2024chr, Cafaro:2024lre, Liu:2025fil} within the model originally formulated in \cite{Kelly:2020lec, Husain:2021ojz, Husain2022, Giesel:2023hys}. \\
\emph{Right diagram}. We propose the following dynamical extension of the LTB solutions. Starting from the trajectories of the LTB shells, we first identify the causally earliest shell-crossing singularity. To the future of this event, we investigate the existence of a junction surface that separates the families of LTB shells such that the induced metric remains continuous across the surface and the surface acts as a boundary between distinct shell evolutions. The shells either emanate from or terminate at this junction surface. Critically, the proper time $t_\pm$ on the LTB shells is continuous prior to shock formation, i.e. $t_+ = t_-$. After the shock forms, however, this proper time becomes discontinuous across the junction surface, as indicated in the plot. Since no further shell crossings occur beyond this point, the spacetime contains no additional SCSs, strictly speaking, apart from the first one. In general, it is possible to find a junction surface with a continuous induced metric, while the extrinsic curvature is discontinuous across it. This discontinuity signals the presence of a thin shell localized on the junction surface. Importantly, the depicted trajectories of each LTB shell are fully determined by the initial data, similarly as in the case illustrated by the left diagram. However, in the right diagram, the junction conditions determine which segments of these trajectories are physically realized at a given time. This {picture} serves as a precursor to a more detailed and fully consistent analysis of the initial-value problem in gravitational theories that are prone to the development of shell-crossing singularities.}
    \label{schematic_plot}
\end{figure*}

In this work, we overcome the limitations of the aforementioned approaches by dynamically imposing the Darmois-Israel junction conditions. Specifically, identifying the first shell-crossing singularity as the origin of the shock, we investigate the existence of a junction surface in its future, approached by dust worldlines that would otherwise intersect, across which the induced metrics are continuous. Such a surface necessarily carries a non-isolated thin shell, as justified below. In this {dynamical extension}, the junction surface acts as a boundary separating two families of dust worldlines (interior and exterior to the shock), such that worldline crossings are described as a regular exchange of matter between the {junction surface} and the surrounding spacetime. The shells do not undergo crossing in the sense of an SCS; instead, the junction conditions force them to terminate at, or emerge from, the boundary. Consequently, successive shell-crossing singularities are systematically replaced by dust lines being absorbed by or emanating from the junction surface. In this scenario, while the continuity of the induced metric is guaranteed by construction, the extrinsic curvature remains discontinuous across the {junction surface} (indicating the presence of a gravitational shock). This discontinuity is proportional to a {non-zero} surface energy-momentum tensor, {therefore the (non-isolated) thin-shell must be present there} \cite{Poisson2004Toolkit}. {This proposal for the dynamical extension beyond shell-crossing singularities is summarized in Fig. \ref{schematic_plot}. }

While the aforementioned picture serves as the conceptual foundation of this study, a simplified version is adopted here due to the computational complexity of its full application. Specifically, we model the formation of the SCS immediately after the bounce as a non-isolated thin shell, expanding as a boundary between an interior—represented by the effective LQC post-bounce solution—and the effective vacuum exterior \cite{Kelly:2020uwj}. This simplification stems from the fact that, in a realistic scenario, the stellar core within the shock is not perfectly homogeneous at the moment of SCS formation, and the exterior is not a perfect vacuum, as a density tail would likely be present and eventually absorbed by the shock. A detailed discussion on the validity of these assumptions is provided in Sec. \ref{sec_modelling}.

The defining feature of the resulting dynamics is that the shell—which is permitted to exchange mass with the interior—is subluminal at all times by construction, and the induced metric remains continuous across the shell. This results in the shell moving toward a second asymptotic vacuum region (distinct from the one containing the black hole horizon) via a white hole region, characterizing inter-universe dynamics. This dynamics is conceptually analogous to the effective Oppenheimer-Snyder scenario {\cite{Bobula:2023kbo, Lewandowski:2022zce, Fazzini:2023scu}} and the isolated thin-shell case \cite{Fazzini25}, involving however a continuous exchange of matter between the interior and the thin shell throughout the evolution.

Although the resulting dynamics may be considered a toy model for spacetime extension beyond shell-crossing singularities—given the simplified interior and exterior solutions—the framework developed herein is applicable to arbitrary solutions. Furthermore, it can be incorporated into the initial value problem {(IVP)} of an effective Lemaître-Tolman-Bondi (LTB) collapse. The present work constitutes a first step in this direction. Finally, it is worth noting that this approach can, in principle, be employed in any classical or modified stellar collapse model that develops shell-crossing singularities.

\section{Effective stellar collapse model and shell-crossing singularities} \label{sec_effe}

As anticipated in the introduction, in this work we consider an effective model of spherically symmetric dust collapse inspired by loop quantum gravity. The model was originally developed in \cite{Husain2022}, where a spatial areal gauge fixing was imposed at the classical level, and subsequently generalized in \cite{Giesel:2023hys}. In the latter, the diffeomorphism constraint is fixed at the effective level, even though it retains its classical phase space dependence. For the purposes of this section, we briefly review the properties of the model in the LTB (or comoving) gauge \cite{Giesel:2023hys}. In this framework, the time coordinate corresponds to the dust proper time, while the radial coordinate remains constant along the dust flow lines, thereby labeling the matter layers of the distribution. The effective metric in this gauge reads
\begin{equation}
 \dd s^2=-\dd t^2 + \frac{[\partial_x r(x,t)]^2}{1+\varepsilon(x)} \dd x^2 + r(x,t)^2 \dd \Omega^2   ~,
\end{equation}
where the time-independent function $\varepsilon(x)$ is set to zero. This function retains its classical interpretation, representing the total (conserved) Newtonian energy of each physical $x$-shell of matter. Setting it to zero (the marginally bound case) implies that every shell composing the star has zero kinetic energy in its asymptotic past ($r(x,t\rightarrow -\infty)\rightarrow +\infty$).
The effective equation of motion, which in these coordinates governs the evolution of the areal radius $r(x,t)$ for the $x$-matter layer, reads
\begin{equation}
\left(\frac{\Dot{r}}{r}\right)^2=\frac{2G M(x)}{r^3}\left(1-\frac{2\gamma^2 \Delta G M(x) }{r^3}\right) ~, \label{eom}
\end{equation}
where {$\Delta = 4 \sqrt{3} \gamma\pi l_P^2$} is the minimum non-zero eigenvalue of the area operator in the full theory \cite{Rovelli:2004tv, Ashtekar:2021kfp}, {and $\gamma$ is the Barbero-Immirzi parameter}. $M(x)$ is the mass function, related to the dust energy density by the following

\begin{equation}
 M(x)=4 \pi \int_0^x \rho(\Tilde{x},t) r^2 (\partial_{\Tilde{x}} r) \dd \Tilde{x}   \label{mass}
\end{equation}
and is a constant of motion for each shell $x$, representing the mass enclosed within that shell. Eq. \eqref{eom} consists of an infinite set of decoupled ODEs, implying that each layer of the distribution evolves independently. However, as in the classical case \cite{Szekeres:1995gy}, the model breaks down upon the formation of shell-crossing singularities. The condition for a shell-crossing singularity is given, analogously to the classical scenario, by

\begin{equation}
\label{conditionscs}
   \mathrm{SCS}:  \begin{cases}
       & \partial_x r(x,t)=0 \\
        & \partial_x M(x) \neq 0 ~.
    
    \end{cases}
\end{equation}
Indeed, by inverting \eqref{mass}

\begin{equation}
 \rho(x,t)=\frac{\partial_x M(x)}{4 \pi r^2 \partial_x r}   ~,
\end{equation}
it is clear that the energy density diverges when \eqref{conditionscs} is fulfilled. Notice that, in contrast to the classical case where $r(x,t)\rightarrow 0$ at the central singularity, here shell-crossing singularity represents the only possible source of divergence for the energy density. This is due to the fact that the shell areal radii---obtained as solutions to the effective equation \eqref{eom}---possess a non-zero minimum: $r_{\mathrm{min}}(x,t_{\mathrm{bounce}})=[ 2G M(x)\gamma^2 \Delta]^{\frac{1}{3}}$. As anticipated in the introduction, the physical interpretation of this kind of singularity is clear: in the spacetime region filled by ordinary matter ($\partial_x M(x) \neq 0$), when two distinct shells of the distribution reach the same areal radius (i.e., when $\partial_x r(x,t)=0$), the volumetric energy density $\rho$ diverges. This corresponds to the formation of an infinitesimally thin matter layer with finite mass. {See Fig. \ref{schematic_plot} for the illustration.}

It has been proved for the dust case---both for marginally bound \cite{Fazzini:2023ova} and non-marginally bound \cite{Cipriani:2024nhx} models---as well as in the presence of pressure \cite{Cafaro:2024lre}, that shell-crossing singularities are unavoidable for any realistic stellar initial density profile. Furthermore, assuming they form for the $x$-shell in the post-bounce phase, always true for initial decreasing density profiles of compact support \cite{Fazzini:2025ysd}, the formation time is $t_\mathrm{SCS}<t_{\mathrm{bounce}}(x)+\frac{2}{3} {\gamma}\sqrt{\Delta}$; thus, they occur within approximately a Planckian time from the bounce of the $x$-shell. {Figure 3 of \cite{Bobula:2024chr} presents a simulation in which this phenomenon is clearly manifested. See also the left diagram in Fig. \ref{schematic_plot} for the schematic plot.}

The previous results show that shell-crossing singularity formation is a central feature of the model, and motivates to seek for a physically sound extension of the spacetime beyond it. This will be the aim of next sections.

\section{Modelling the non-isolated thin shell as the future of SCS} \label{sec_modelling}

{The detailed implementation of the general strategy for obtaining a consistent evolution of stellar collapse, as outlined in the right diagram of Fig. \ref{schematic_plot}, becomes computationally demanding for more realistic initial dust profiles (e.g., Gaussian distributions, {see the left diagram in Fig. \ref{conformal}}). For this reason, we restrict our analysis to a simplified scenario that nevertheless preserves the essential ideas and overall framework illustrated in Fig. \ref{schematic_plot}.} {Specifically, in this work, we fill formulate a toy model for the future of the first SCS, as depicted in the right diagram Fig. \ref{schematic_plot}. }

The Darmois-Israel junction conditions, which generally allow for the study of the evolution of a spacetime where different solutions of the field equations are matched at a common boundary, are typically applied in simplified contexts, such as the Oppenheimer-Snyder model or the isolated thin-shell case \cite{Poisson2004Toolkit}. Regarding the effective model under consideration, several works have explored both directions (see \cite{Bobula:2023kbo,Lewandowski:2022zce,Fazzini:2023scu,Giesel:2023hys} for the effective Oppenheimer-Snyder case, and \cite{Fazzini25} for thin-shell collapse). The scenario addressed here is more complex, as the (singular) boundary we investigate through the Darmois-Israel formalism forms \emph{during} the dynamics (at the SCS) from regular initial data. Moreover, this is not an isolated thin shell; when the first SCS occurs during the LTB evolution, the region inside the shell is filled with matter and is, in general, dynamical. Conversely, in the isolated thin-shell case, the interior is typically assumed to be Minkowski and the exterior an effective Schwarzschild metric at all times \cite{Fazzini25}.

In more detail, we generalize the formulation presented in \cite{Fazzini25}. Specifically, we consider an effective Schwarzschild metric for the exterior, while for the interior we adopt the effective solution from \cite{Kelly:2020lec}, which constitutes the spherically symmetric formulation of the effective LQC solution \cite{Ashtekar:2006wn}.  We stress at this point that these requirement does not mean the study of the effective Oppenheimer-Snyder collapse. Even though we model the interior as an homogeneous cosmology, and the exterior as vacuum, we allow a thin dust shell to be present at the boundary (due to SCS formation), which requires the violation of the second Darmois-Israel junction condition, which becomes an evolution equation for the shell itself. 
{This scenario is related to the Fig. \ref{schematic_plot} as follows: here the shells bouncing at early times encode homogeneous and isotropic geometry whereas the shells bouncing at late times encode vacuum geometry.} We are therefore implicitly assuming that the density profile before the bounce has a negligible tail, ensuring that few matter shells intersect during the bounce of the star.  Indeed, we set the initial conditions in the post-bounce phase for the interior and describe the exterior using the effective Schwarzschild metric in retarded PG coordinates, which capture the non-trapped region, the white hole region, and the second asymptotic region of the maximal extension of the effective Schwarzschild spacetime \cite{Kelly:2020uwj}. Furthermore, we assume a positive shell radial velocity at the initial time (representing an expanding non-isolated thin shell) and locate the shell between $r_\mathrm{min}$
  and the inner horizon of the effective vacuum solution \cite{Lewandowski:2022zce}.

While these constraints on initial conditions are restrictive, they provide a viable toy model for the general picture: a realistic collapsing star consists of an almost homogeneous interior with a density profile tail expected to be quite sharp, modelling the stellar atmosphere. Unlike the Oppenheimer-Snyder case, this type of profile develops shell-crossing singularities after the bounce \cite{Fazzini:2023ova}, although the actual number of crossing shells may be small, depending on the sharpness of the tail. Consequently, for such profiles, the gravitational mass of the resulting thin shell would be small compared to the interior mass. For profiles with an almost homogeneous and isotropic interior, the dynamics is expected to preserve these symmetries throughout the spacetime evolution, even beyond shell-crossing singularity formation; this justifies assuming an exact FLRW dynamics for the interior of our non-isolated thin shell. Since the mass in the tail undergoing shell-crossing is small, we assume that shortly after the first crossing, the entire mass of the tail is incorporated into the gravitational mass of the thin shell. The exterior, left empty after the tail accumulates on the shell, is then described by the effective Schwarzschild metric.

Finally, the relative velocity between the internal fluid lines and the shell is assumed to be large and negative at the initial time. This is because the shell is formed by the crossing of expanding internal fluid lines and collapsing lines from the tail—a process generally characterized by high impact velocities.

\section{Darmois-Israel junction conditions} \label{sec_darmois}

The formulation of the Darmois-Israel junction conditions will be carried out in PG coordinates. This choice is not restrictive, as the two spacetimes joined via the junction conditions at the thin shell do not need to be expressed in specific coordinate systems. One could, in this context, use two different LTB charts for the interior and exterior, as is done, for example, in the classical study of cosmological voids \cite{Maeda1983}.

The idea, therefore, is to consider as the interior an effective FLRW spatially flat—or, in LTB terms, marginally bound—solution, whose metric in Painlevé-Gullstrand coordinates reads
\begin{equation} \label{FLRW_PG}
 \dd s_-^2=-\dd t_-^2 + (\dd r_-+ N^r_- \dd t_-)^2 + r_-^2 \dd \Omega^2  ~, 
\end{equation}
where 
\begin{equation} \label{shift_frw}
  N^r_-=-\frac{6 r_- t_-}{9 t_-^2+4\gamma^2 \Delta}  
\end{equation}
is the radial component of the shift vector
\cite{Husain2022}, and {$t_-=t_\mathrm{bounce} =0$} at the bounce. The exterior instead is given by the effective Schwarzschild metric in retarded PG coordinates
\begin{align}
 \dd s^2=&-\left(1-\frac{R_S}{r_+}+\frac{\gamma^2 \Delta R_S^2}{r_+^4} \right) \dd t_+^2 - \notag \\
 & -2 \sqrt{\frac{R_S}{r_+}\left( 1-\frac{\gamma^2 \Delta R_S}{r_+^3}\right)  } \dd t_+ \dd r_+ + \dd r_+^2 +r_+^2 \dd \Omega^2
 \label{pg+}
\end{align}
where $R_S=2GM$ is the classical Schwarzschild radius and $M$ is the Schwarzschild mass parameter, that coincides with the ADM mass of the system. For masses larger than the critical one ($m_\mathrm{crit.}\sim 8\gamma \sqrt{\Delta}/\sqrt{27}G\sim m_\mathrm{Planck}$), there are two null horizons, inner and outer as in the classical Reissner-Nördstrom solution, located at 
\begin{align} \label{horizon_radii}
&r_{\mathrm{in}}= (\gamma^2 \Delta R_S)^{\frac{1}{3}}+ \frac{1}{3}\left( \frac{\gamma^4 \Delta^2}{R_S}\right)^{\frac{1}{3}}+O\left(\frac{\Delta^\frac{4}{3}}{R_S^{\frac{5}{3}}} \right) ~,  \\
&r_{\mathrm{out}}=R_S-\frac{\gamma^2 \Delta}{R_S}+O\left(\frac{\Delta^\frac{4}{3}}{R_S^\frac{5}{3}} \right)~.
\end{align}
As one can easily notice from the sign of the off-diagonal term, which determines the shift vector
\begin{equation}
\label{shift+}
   N^r_+=- \sqrt{\frac{R_S}{r_+}\left( 1-\frac{\gamma^2 \Delta R_S}{r_+^3}\right)  } \,,
   \end{equation}
in the exterior metric, this is the time-reverse of the effective black hole solution presented in \cite{Kelly:2020uwj}, and contains an anti-trapped region. Notice that, as in the classical Reissner-Nordström case, this solution can be viewed as a part of the maximal extension of the vacuum solution \cite{Lewandowski:2022zce,Fazzini:2023scu}.

The first condition to be imposed is the continuity of the metric across the thin shell. To this end, we first express the induced metric on the thin shell as measured from both the interior and the exterior
\begin{equation}
\dd s_\pm^2=\left[\left((N^r_\pm)^2-1 \right)\Dot{T}_\pm^2+2N^r_\pm \Dot{T}_\pm \Dot{R}_\pm+\Dot{R}_\pm^2\right]\dd \tau^2 + R_\pm^2 \dd \Omega^2 ~,
\end{equation}
where $\tau$ is the shell proper time, the overdot is the derivative with respect to $\tau$, $T_\pm\equiv t_\pm (\tau)$, $R_\pm\equiv r_\pm(\tau)$. Continuity of the induced metric requires (first Darmois-Israel junction condition) $\dd s^2_+=\dd s^2_-$, that implies $\theta_+=\theta_-$, $\phi_+=\phi_-$, $R_+=R_-$, and $g_{\tau \tau}|_{\Sigma}$ being the same as computed with the interior and exterior metric. Notice that requiring $\tau$ being the shell proper time implies a further restriction on the induced metric
\begin{equation}
\label{firstisrael}
 \left(1-(N^r_\pm)^2 \right)\Dot{T}_\pm^2-2N^r_\pm \Dot{T}_\pm \Dot{R}-\Dot{R}^2 =1~, 
\end{equation}
which automatically ensures the timelike character of the thin shell, as measured from \emph{both} the exterior and interior metrics, and is compatible with the continuity condition. 
To check this, one can simply compute the norm squared of the thin shell 4-velocities $u^\mu_\pm=\left\{{\frac{\dd T_\pm}{\dd \tau},\frac{\dd R}{\dd \tau}, 0,0  } \right\}$. 

For later convenience, we construct the space-like unit normal $4$-vector, which is orthogonal both to the thin-shell spatial surface and the thin shell $4$-velocity $u^\mu_\pm$:

\begin{align}
 \chi^\mu_\pm 
 = \begin{pmatrix}
\frac{\Dot{R}+ N^r_\pm \sqrt{\Dot{R}^2+1-(N^r_\pm)^2}}{1-(N^r_\pm)^2} \\
\\
 \sqrt{\Dot{R}^2+1-(N^r_\pm)^2}\\
0\\
0    \label{chipm} 
\end{pmatrix} ~,
\end{align}
derived by requiring $\chi^\mu_\pm \chi_\mu^\pm=1$, $\chi^\mu_\pm u_\mu^\pm=0$, $\chi^\theta=\chi^\phi=0$.

Condition \eqref{firstisrael} can be rearranged to 

\begin{equation}
 \label{dotsT}
  \Dot{T}_\pm=\frac{\Dot{R} N^r_{\pm}+\sqrt{\Dot{R}^2+1-(N^r_\pm)^2}  }{1-(N^r_\pm)^2}  ~.
\end{equation}
The strength of this first condition lies in the fact that we do not need to verify whether the induced metric on the shell hypersurface changes signature during the dynamics, as its preservation—which is necessary from a physical point of view—is automatically ensured.

Next, we impose the second Darmois-Israel junction condition (or rather, the junction condition for a thin shell with a surface stress-energy tensor) \cite{Poisson2004Toolkit}. We follow a Hamiltonian formulation, similar to the isolated thin-shell case \cite{Fazzini25}, as the LTB solutions for both the interior and exterior are derived from the Hamiltonian framework in Ashtekar-Barbero variables, characteristic of LQG-inspired models. The idea behind this formulation of the second Darmois-Israel condition is to begin by imposing the spatially integrated version of the classical scalar and diffeomorphism constraints across the thin shell, which is generally located at $r=R$
\begin{align}
  &   \int_{R-\varepsilon}^{R+\varepsilon} N(\mathcal{H}^\mathrm{grav.}_t+ \mathcal{H}^\mathrm{matt.}_t+\mathcal{H}^\mathrm{shell}_t   )\dd r \approx 0 \notag ~,\\
& \int_{R-\varepsilon}^{R+\varepsilon} N^r(\mathcal{H}^\mathrm{grav.}_r+\mathcal{H}^\mathrm{matt.}_r +\mathcal{H}^\mathrm{shell}_r   )\dd r\approx 0  ~, \label{diffeo}
\end{align}
with $\varepsilon/R \ll 1$,
and, after spatial gauge fixing, polymerize the resulting shift vector and gravitational part of the Hamiltonian constraint, through the so called $\bar{\mu}+K$-loop quantization scheme \cite{Husain2022}, typical of the effective model under consideration. Then, impose the integral version of the resulting Hamiltonian constraint across the thin shell surface to vanish, and eventually take the limit $\varepsilon \rightarrow 0$. The procedure is carried out explicitly in appendix \ref{appendix0}.
The resulting equation reads
\begin{equation}
 M-\frac{4}{3}\pi \rho R^3=-\lim_{\varepsilon\rightarrow 0} \int_{R-\varepsilon}^{R+\varepsilon}\dd r (N^r_{\pm} {^\pm\mathcal{H}}^\mathrm{shell}_r+{^\pm\mathcal{H}}^\mathrm{shell}_t) ~,   \label{finale}
\end{equation}
where $M$ is the Schwarzschild mass for the exterior \eqref{pg+}, $\rho$ is the dust energy density of the FLRW interior---which is solution of the effective Friedmann equation \cite{Husain2022}---and $N^r_+$ is given by \eqref{shift+} (valid for $r\geq R$) while $N^r_-$ by \eqref{shift_frw} (valid for $r< R$). The thin-shell contributions to the Hamiltonian and diffeomorphism constraints are instead given by \cite{Fazzini25,Crisostomo:2003bb}
\begin{align}
 &{^\pm\mathcal{H}}^\mathrm{shell}_t= 4 \pi R^2(n \cdot u)_\pm^2 \sigma \delta(\chi_\pm) ~,\\
 & {^\pm\mathcal{H}}^\mathrm{shell}_{r}=4 \pi R^2 (n \cdot u)_\pm u^\pm_r \sigma \delta (\chi_\pm) ~, \label{lastof}
  \end{align}
where the $+$ ($-$) holds for $r\geq R$ ($r<R$). $n^\mu$ in the previous expressions is the $4$-vector aligned with the PG flow, i.e., $n^\mu_-=u^\mu_F$ (the FLRW fluid $4$-velocity), $n^\mu_+=\left( 1,- N^r_+,0,0  \right)$. Furthermore, $u_\pm^\mu$ is the thin-shell $4$-velocity as measured from the exterior ($+$) and interior ($-$) metrics, $\sigma$ is the proper surface density of the shell, and $\chi_\pm$ is the radial coordinate along $\chi_\pm^\mu$ \cite{Fazzini25}. Note that, in general, $u^\mu_F$ differs from $u_-^\mu$, as the thin shell does not necessarily follow an FLRW geodesic, despite being composed of dust. 
The explicit computation of the thin-shell contribution to the constraints was carried in \cite{Fazzini25} for the isolated thin-shell case. Following the same procedure, one gets
\begin{align}
&{^\pm\mathcal{H}}^\mathrm{shell}_t=4 \pi \sigma R^2 \frac{\Dot{R}N^r_\pm+\sqrt{\Dot{R}^2+1- (N^r_\pm)^2} }{1-(N^r_\pm)^2} \delta (r-R) ~, \\
&{^\pm\mathcal{H}}^\mathrm{shell}_r=-2\pi R^2 \sigma \frac{\Dot{R}+N^r_\pm \sqrt{\Dot{R}^2+1-(N^r_\pm)^2}  }{1-(N^r_\pm)^2}\delta(r-R) ~.
\end{align}
 By integrating the RHS of \eqref{finale}, writing everything explicitly, 
\begin{align}
 M-\frac{4}{3}\pi \rho(T_-) R^3=&
2 \pi \sigma R^2 \bigg( \sqrt{\Dot{R}^2+1-\frac{R_S}{R}+\frac{\gamma^2 \Delta R_S^2}{R^4}} + \notag \\
 &+\sqrt{\Dot{R}^2+1-\frac{36 R^2 T_-^2}{(9 T_-^2+4 \gamma^2 \Delta)^2  }  }  \bigg) ~,
\label{shelleq}
\end{align}
which gives the correct classical limit for $R \gg (\gamma^2 \Delta R_S)^{1/3}$, $\rho\ll \rho_\mathrm{crit.}$ \cite{Maeda1983,Sakai1993,Sakai:1999xx}. Furthermore, notice that if the interior is Minkowski, we recover the equation for the effective isolated thin-shell evolution \cite{Fazzini25}. Equation \eqref{shelleq} can be interpreted as an energy balance for the shell; specifically, it states that the total mass of the system (the Schwarzschild mass) minus the mass of the interior equals the total energy of the shell (RHS). However, this equation alone is insufficient to fully determine the thin-shell motion, since depends on the dynamical variables $\sigma(\tau)$ and $R(\tau)$. It must be supplemented by an equation describing the evolution of the shell's mass during its dynamics. In the case of isolated thin-shell collapse, the gravitational mass of the shell remains constant, as does its rest mass $m=4\pi R^2 \sigma$ \cite{Fazzini25}, and the evolution of $\sigma$ is uniquely determined by $R(\tau)$. In the non-isolated case, since both the interior and the thin shell are composed of dust, it is reasonable to assume that they can exchange gravitational (and inertial) mass. Moreover, since the relative velocity between the interior dust particles and the thin shell is generally non-zero, an energy exchange will occur. The equation governing this process is provided by the covariant energy-momentum conservation law across the shell.
It reads (see appendix \ref{appendix1} for the derivation)
\begin{equation}
 \frac{\dd}{\dd \tau}(4 \pi R^2 \sigma)=4 \pi R^2 [(T^{\mu \nu}_+ \chi^+_\mu u^+_\nu  )|_\Sigma- (T^{\mu \nu}_- \chi^-_\mu u^-_\nu)|_\Sigma  ]   ~,
\end{equation}
where $T^{\mu \nu}$ is the energy momentum tensor of the exterior $(+)$ and interior $(-)$. The exterior, being the effective Schwarzschild spacetime, is described by $T_+^{\mu \nu}=0$, while for the interior we have
\begin{equation}
    T_-^{\mu \nu}=\rho u_F^\mu u_F^\nu ~,
\end{equation}
where $\rho$ is the interior FLRW dust energy density. We can then write the conservation more explicitly as
\begin{equation}
\frac{\dd }{\dd \tau} (4 \pi R^2 \sigma)=-4 \pi R^2 \rho u_F^\mu  \chi^-_\mu u_F^\nu u^-_\nu~.
\label{consmass}
\end{equation}

Notice that, as previously anticipated, the non-vanishing matter flux across the shell (i.e. the non-zero right-hand side) arises from the misalignment between the dust fluid $u_F^\mu$ and the shell $4$-velocity. This misalignment results in a non-vanishing scalar product between the fluid 4-velocity and the $4$-vector $\chi^\mu$, since $u_-^\mu \chi^-_\mu=0$. Physically, this represents a net flux of dust particle toward the shell surface, leading to a variation in its proper (and consequently gravitational) mass. Were the two $4$-velocities aligned, this flux would vanish.

By assuming a vacuum exterior, we are requiring that matter can be exchanged between the shell and the interior, but not with the exterior. While this is not the most general case, it is consistent with the assumption of a sharp stellar density tail. A generalization to configurations featuring dust matter in the exterior region is left for future work. A direct computation of the conservation of the energy-momentum tensor leads to the following evolution equation for the shell mass
\begin{equation}
\frac{\dd }{\dd \tau} (4 \pi R^2 \sigma)=-4 \pi R^2 \rho \Dot{T}_- (\Dot{R} +\Dot{T}_- N^r_-  )~.
\end{equation}
This equation governs the exchange of mass between the interior and the thin shell at all times and is non-trivial. Specifically, if we compute the variation of the gravitational mass within the FLRW region, defined as $M_\mathrm{FLRW}=\frac{4}{3}\pi \rho R^3$ where $R(\tau)$ denotes the shell's location, we obtain
\begin{equation}
  \frac{\dd M_\mathrm{FLRW}}{\dd \tau}=-\frac{1}{\Dot{T}_-}\frac{\dd }{\dd \tau}  (4 \pi R^2 \sigma) ~.
  \label{finalconse}
\end{equation}
 To better understand this expression, we can rearrange Eq. \eqref{dotsT} as follows:
\begin{equation}
    \Dot{T}_-=\frac{1}{\sqrt{1- \left(N_-^r+\frac{\dd R}{\dd T_-}\right)^2}}~,
\end{equation}
where $\dd R / \dd T_-$ is the radial coordinate velocity of the shell as measured from the FLRW interior, and $-N^r_-$ is the proper (but also coordinate) radial velocity of a dust geodesic particle in the FLRW spacetime. Consequently, the term $ N_-^r+\dd R/ \dd T_-$ is the relative coordinate velocity between the dust flow and the shell (sometimes referred to in the literature as the peculiar velocity of the shell \cite{Sakai:1999xx}).

In this context, $\Dot{T}_-$ can be interpreted as the instantaneous Lorentz factor $\Gamma$ between the frame comoving with the interior, and the shell's rest frame. Therefore, the relationship between the variation of the two masses reads
\begin{equation}
    \dd m =- \Gamma \dd M_\mathrm{FLRW} ~.
\end{equation}
This implies that when the FLRW fluid loses an amount of mass $\dd M_\mathrm{FLRW}$, the shell acquires a total mass corresponding to the relativistic energy associated with that mass. Due to the relative motion, the shell gains not only the rest mass $\dd M_{FLRW}$ but also the kinetic energy $(\Gamma-1) \dd M_{FLRW}$ transferred during the accretion process.

To conclude this section, we can summarize the dynamical evolution of the system. The numerical or analytical resolution of the model requires solving a coupled system of ordinary differential equations that describes both the trajectory of the shell and the evolution of its mass
\begin{equation} \label{newFRW}
\begin{aligned}
& M-\frac{4}{3}\pi \rho(T_-) R^3=2 \pi \sigma R^2 \left( \sqrt{\dot{R}^2+F_+} +\sqrt{\dot{R}^2+F_-}  \right)~,\\
 &\frac{\dd }{\dd \tau} ( R^2 \sigma)=- R^2 \rho(T_-) \dot{T}_- (\dot{R} +\dot{T}_- N^r_-  )~, \\
   &\dot{T}_-=\frac{\dot{R} N^r_{-}+ \sqrt{\dot{R}^2+F_-}  }{F_-}  ~,
 \end{aligned}
\end{equation} 
where we have introduced $F_\pm\equiv 1-(N^r_\pm)^2$.
We can further simplify Eq. \eqref{newFRW} into a system of two coupled equations evolving with respect to the interior coordinate time $T_-$, since the PG chart for the interior covers the whole shell evolution (as well as the whole interior evolution). To do this, we first algebraically solve for $\dot{R}^2$ using the first equation in \eqref{newFRW} 
\begin{equation} \label{uu}
\begin{aligned}
  u\equiv \dot{R}^2=  &-\frac{F_+}{2} - \frac{F_-}{2} + \frac{(3 M - 4 \pi R^3 \rho)^2}{144 \pi^2 R^4 \sigma^2}+ \\&+ \frac{9 (F_+ - F_-)^2 \pi^2 R^4 \sigma^2}{(3 M - 4 \pi R^3 \rho)^2} ~.
\end{aligned}
\end{equation}
Then, we use relations $R' \equiv \frac{\partial R}{\partial T_-} = \dot{R}/\dot{T}_-$ and $\sigma'\equiv \frac{\partial \sigma}{\partial T_-} = \dot{\sigma}/\dot{T}$ to write
\begin{equation}  \label{newcoupledFRW}
\begin{aligned}
\sigma' &=   - \frac{2 \sigma  R'}{R} -  \rho \left( N_- + R'\right) \frac{\mathrm{sgn}(R') \sqrt{u} N^r_{-}+ \sqrt{u+F_-}  }{F_-}~, \\
  R' &=  \frac{F_-}{ N^r_{-}+ \mathrm{sgn}(R') \sqrt{ \frac{u+F_-}{u}}  } ~,
\end{aligned}
\end{equation} 
where $\mathrm{sgn}$ denotes a signum function. Thus, \eqref{newcoupledFRW} represents the simplified ODE system that captures the thin-shell dynamics with respect to $T_-$ coordinate time. 
\section{Numerical results for evolution and conformal diagrams}
\subsection{Thin shell dynamics} \label{sec_thin_shelldynamics}

With the physical context and dynamical equations for the scenario under consideration established in Secs. \ref{sec_modelling} and \ref{sec_darmois}, we proceed to the numerical analysis of the non-isolated thin-shell evolution.

    We elaborate on the initial conditions required to solve the system~\eqref{newcoupledFRW}. The initial values of $T_-=T_{\text{ini}}$ and $R=R_{\text{ini}}$ correspond to the spacetime location of the first SCS emerging during the collapse process (more specifically, immediately after the bounce of the stellar core), which also marks the starting point of the non-isolated thin-shell evolution. Once these are fixed, one must choose the initial value of $\sigma$. However, due to Eq. \eqref{uu}, this choice is equivalent to specifying the initial coordinate velocity of the shell, say $\dot{R}=\dot{R}_{\text{ini}}$. As anticipated at the end of Sec. \ref{sec_modelling}, we restrict our analysis to the cases where the initial relative velocity $N^r_-+R'$ (see Sec. \ref{sec_darmois}) is negative and large at the initial $(T_\text{ini},R_\text{ini})$ point. Such restriction considerably reduces the span of possible values of the initial $\dot{R}$, however it does not ultimately fix it. Within the simplified scenario of SCS formation of our consideration we are not able to find a fully unambiguous resolution for this issue, which would be avoided if this approach is included in an IVP formulation. 

We will also compute the PG time discontinuity at the non-isolated thin shell. We can combine Eqs. \eqref{dotsT} \eqref{uu} together to get
\begin{equation} \label{TplusTminus}
\frac{d T_{+}}{d T_{-}}=\frac{F_{-}}{F_{+}} \left(\frac{\operatorname{sgn}\left(R^{\prime}\right) \sqrt{u} N_{+}^r+\sqrt{u+F_{+}}}{\operatorname{sgn}\left(R^{\prime}\right) \sqrt{u} N_{-}^r+\sqrt{u+F_{-}}} \right) \,.
\end{equation}
The above function quantifies the time discontinuity across the non-isolated thin shell. Since it is generally not constant, it implies that the time coordinates $T_-$ and $T_+$ evolve at different rates on the two sides of the shell, differently from what happens in approaches based on weak solutions \cite{Husain2022,Fazzini:2025hsf,Liu:2025fil}.

\begin{figure*}
    \centering
\includegraphics[width=0.85\linewidth]{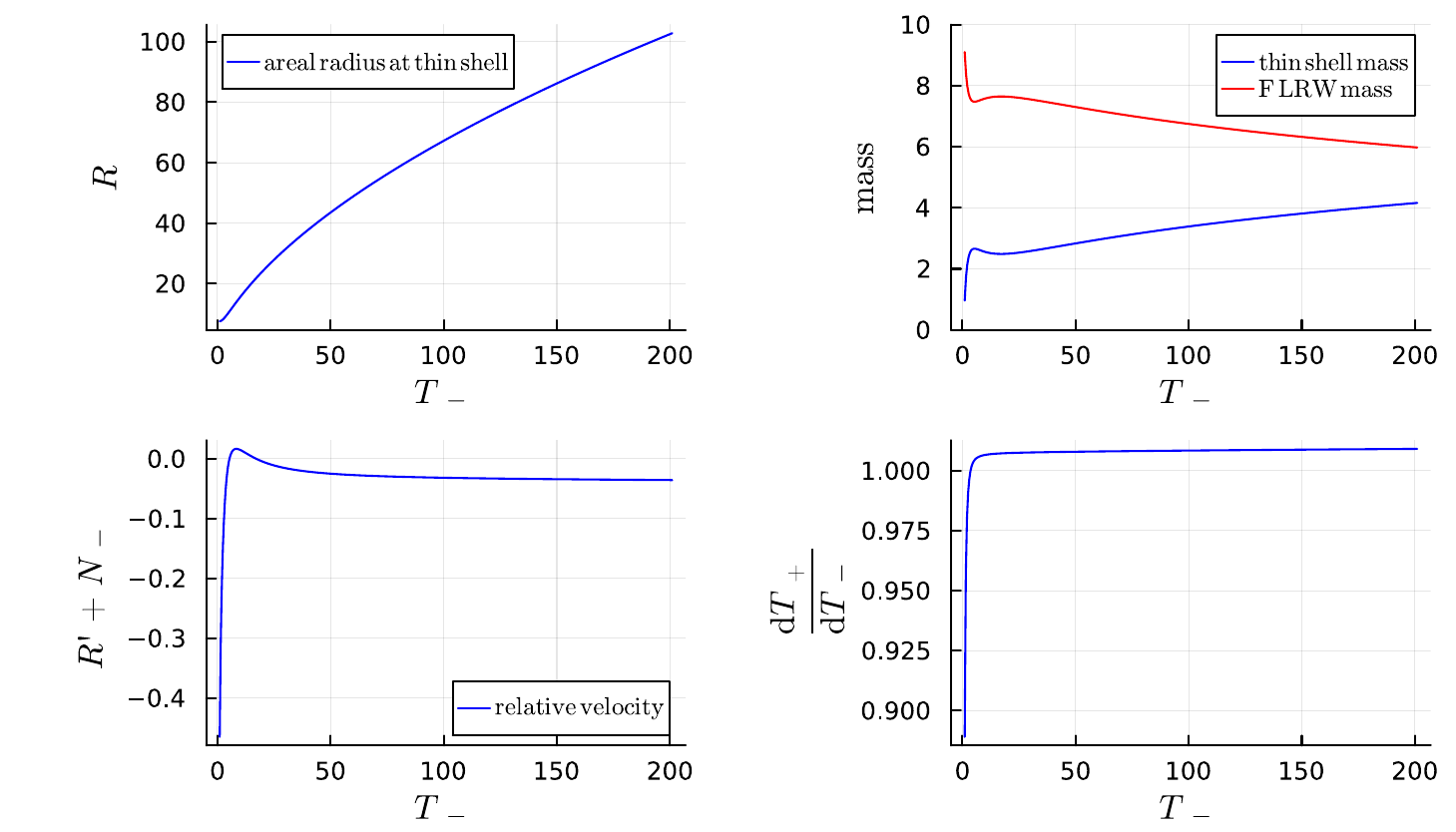}
    \caption{Numerical results for the non-isolated thin shell as a future of a shell crossing singularity. \emph{Top left.} Time evolution of areal radius at the non-isolated thin shell.  \emph{Top right.} The time evolution of the FLRW mass {$M_\mathrm{FLRW}$} along with the inertial thin shell mass $4 \pi \sigma R^2$. 
    \emph{Bottom left.} The coordinate relative velocity of the thin shell with respect to the FLRW interior. \emph{Bottom right.} The discontinuity between the time coordinates $T_-$ and $T_+$ across the non-isolated thin shell characterized by the function $\mathrm{d}T_-/\mathrm{d}T_+$. Since this function is not constant, the two time coordinates do not evolve uniformly.
    For each plot we have taken $M=10$, $\gamma=1$, $T_{\text{ini}}=1$, $R_\text{ini}=r_\text{min}+0.01$,  where $r_\text{min}=r_\text{min}(x=1, t_\mathrm{bounce})$ and $\dot{R}_\mathrm{ini}=0.01$ in Planck units. }
    \label{3plots}
\end{figure*}

Fig. \ref{3plots} displays the numerical results for the non-isolated thin shell, obtained using libraries in \texttt{Julia} programming language. For the chosen initial conditions, the thin shell expands in areal radius, while its mass generally increases at the expense of the interior FLRW mass. Indeed, the small initial shell mass becomes large and macroscopic at late times. The initially large and negative velocity rapidly evolves at the beginning and then quickly settles to a small negative value close to zero. A coordinate time discontinuity is observed across the shell. 

The plots do not rule out the possibility that the thin shell persists indefinitely, however, neither our analytical nor numerical analysis provides a definitive answer to this question. Nevertheless, the areal radius of the thin shell becomes significantly larger than the horizon radii given by Eqs. \eqref{horizon_radii} suggesting possible escape to infinity.

\subsection{Conformal diagram}
\begin{figure*}
    \centering
    \includegraphics[width=1\linewidth]{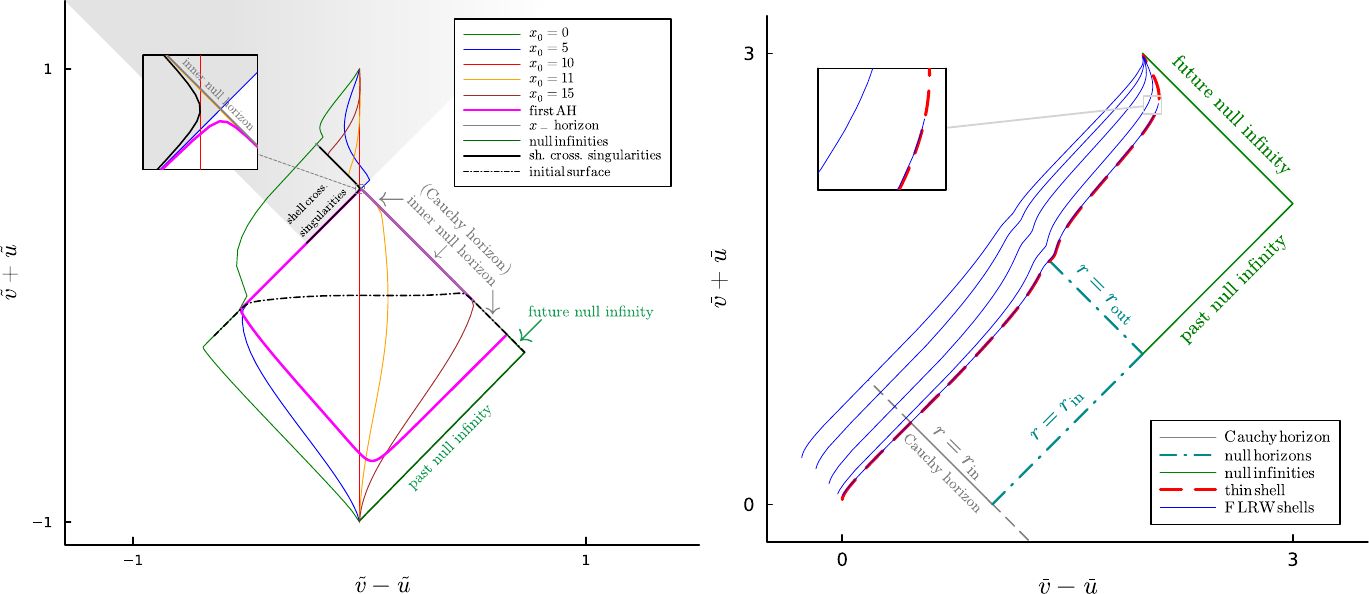}
    \caption{\emph{Left diagram}. The conformal diagram for an inhomogeneous dust collapse with an initial Gaussian density profile, obtained within the analysis of the LQG-inspired model \cite{Kelly:2020lec, Husain:2021ojz, Husain2022, Giesel:2023hys}, presented in \cite{Bobula:2024chr} (see also Figs. 3 and 9 therein). The LTB shells evolve from the initial time surface and subsequently undergo shell crossings in the deep Planck regime. Shell-crossing singularities are therefore present, and the future light cone emanating from the causally first shell-crossing singularity defines the \emph{shadow}—the greyed-out region—in which the employed dynamical extension beyond the first SCS is expected to be unreliable. The spacetime structure in this region constitutes a concrete realization of the schematically depicted dynamical extension shown in the left diagram of Fig. \ref{schematic_plot}. \\
    \emph{Right diagram}. The numerically computed conformal diagram for the toy model analysed in this work illustrates a consistent dynamical extension beyond shell-crossing singularities, compatible with the right schematic plot in Fig.~\ref{schematic_plot}, and covering the grayed-out region in the left diagram. A non-isolated thin shell emerges at the spacetime point where the causally first shell-crossing singularity occurs. In the toy model considered here, we employ an approximation in which the interior LTB shells are treated as homogeneous and isotropic FLRW shells. Furthermore, the shells constituting the tail of the initial Gaussian density profile are assumed to have already been collected into the thin shell at the initial time of the evolution (shown in Fig. \ref{3plots}). The non-isolated thin shell emerges from the deep Planck-regime region, crosses the first inner null white-hole horizon (which also constitutes a Cauchy horizon), then crosses the outer white-hole horizon, and finally approaches the vicinity of timelike infinity in the new asymptotic region. \emph{Inset.} As shown in the right schematic plot in Fig.~\eqref{schematic_plot}, the non-isolated thin shell can either collect or release LTB shells. Here we illustrate a single instance of this process, in which the thin shell collects one of the interior FLRW shells. {This diagram is generated using the same model parameters and initial conditions as in Fig. \ref{3plots}, in addition, we set $\kappa = 0.05$. The interior FLRW shells shown correspond to comoving radii $x \in \{0,0.25,0.5,0.75,0.91\}$. All quantities are expressed in Planck units.}} 
    \label{conformal}
\end{figure*}

We will compute Penrose-Carter conformal diagram for the analysed non-isolated thin shell in Sec. \ref{sec_thin_shelldynamics}. The construction follows the methodology developed in Refs. \cite{Bobula:2024chr, Bobula:2024ywp}. The results are summarized in Fig. \ref{conformal}.

To represent the exterior effective Schwarzschild vacuum geometry on the conformal diagram, we first introduce a pair of double-null coordinates ($v,u$) defined as follows
\begin{equation}
    \begin{aligned}
        v=t_+ +\int^r_{r_\mathrm{min}} \frac{1}{1+N_+^r}\dd r \,, \\ 
        u=t_+ -\int^r_{r_\mathrm{min}} \frac{1}{1-N_+^r}\dd r\,.
    \end{aligned}
\end{equation}
Note that these coordinates satisfy the necessary condition for any system of double-null coordinates \cite{Bobula:2024chr}, namely that their exterior derivatives obey $\dd(\dd v)=0$ and $\dd(\dd u)=0$. In these coordinates, the trajectory of the non-isolated thin shell is described by $(v(T_+, R(T_+)), u(T_+, R(T_+) )$ where the exterior time coordinate $T_+$ can be expressed as a function of the interior time $T_-$, i.e.\ $T_+ = T_+(T_-)$, by integrating Eq.~\eqref{TplusTminus} with an initial condition $T_+(T_\mathrm{ini})=T_+^\mathrm{ini}$. The choice of  $T_+^\mathrm{ini}$ is free in our analysis, as the exterior vacuum metric is time-independent\footnote{It is important to note that $T_+$ is the time coordinate of the thin shell as measured by the external metric $g_+$, and represents the proper time of the congruence of timelike geodesics defining the PG chart in the external effective Schwarzschild region. Fixing the value of $T_+$ at a single spacetime point determines its normalization up to an additive constant at that point, however, the values at all other spacetime points must then be adjusted consistently with this choice.} . Throughout this work, we choose $T^\mathrm{ini}_+ = T_\mathrm{ini}$. 

For the non-isolated thin shell trajectory the null coordinate diverges, with $v\to \infty $ and $v\to -\infty$ as $R\to r_\mathrm{in}$ and $R \to r_\mathrm{out} $ respectively, whereas the $u$ remain finite and monotonic throughout the evolution. So, to cover the whole evolution within single coordinate chart, we incorporate following compactification 
\begin{equation} \label{uvcompactout}
\begin{aligned}
    \bar{v} &= \pm \frac{1}{\pi} \arctan (\kappa v) + c_v\,,  \\
     \bar{u} &= \pm \frac{1}{\pi} \arctan (\kappa u) + c_u\,,
\end{aligned}
\end{equation}
where $c_v \in \{0,1,2\}$ and $c_u = 0$ are diagram's \emph{block} constants \cite{Bobula:2024chr} and $\kappa$ is a dimensionless parameter introduced to adjust the visual appearance of the conformal diagram. In the above compactified expressions, the signs and the constants must be switched whenever $u$ or $v$ diverges, so as to preserve the monotonicity of $\bar{u}$ and $\bar{v}$. Indeed, for the non-isolated thin shell trajectory, the compactified $\bar{v}$ coordinate is constructed as follows. At the origin of the shell one uses the plus sign with $c_v=0$, after crossing the $r_{\mathrm{in}}$ horizon the sign switches to minus with $c_v=1$, finally, once the $r_{\mathrm{out}}$ horizon is crossed, the plus sign is restored with $c_v=2$. By contrast, the compactified $\bar{u}$ coordinate retains the plus sign with $c_u=0$ throughout the entire trajectory.

So far, we have constructed the coordinates~\eqref{uvcompactout} for the effective Schwarzschild vacuum. We now extend these coordinates to the interior FLRW geometry. As a first step, we rewrite the interior metric in terms of the comoving radial coordinate~$x$. Indeed, starting from Eqs.~\eqref{FLRW_PG} and~\eqref{shift_frw}, we introduce the scale factor~$a$ via the relations
\begin{equation} \label{toFLRW}
\begin{aligned}
    N^r_- &=- \frac{\dot{a}(t_-)}{a(t_-)} r_-\,, \\
    r_{-}&=a(t_-) x\,.
\end{aligned}
\end{equation}
With these relations, the metric \eqref{FLRW_PG} takes the familiar FLRW form
\begin{equation}
    \dd s_-^2 = -\dd t_-^2 + a(t_-)^2\,\dd x^2 + a(t_-)^2 x^2\,\dd\Omega^2 \, .
\end{equation}
{Throughout our analysis, the mass defined in Eq.~\eqref{mass} is related to the Schwarzchild mass by $ M(x=1)=M $. This choice fixes the freedom to redefine the comoving coordinate $x$.} The scale factor $a(t_-)$ follows from integrating the first equation in \eqref{toFLRW}. Imposing the initial condition
$a(0)=\left. r_\mathrm{min}/x \right|_{x=1}$,
which is consistent with the {abovementioned choice for the comoving radius}, we obtain
\begin{equation}
    a(t_-)=\left(\frac{9 G M t_{-}^2}{2}+2\gamma^2 \Delta GM\right)^{1 / 3}\,,
\end{equation}
and the interior FLRW shells are labeled by $x\in [0,1]$. Now, we ready to write double null coordinates for the interior as 
\begin{equation} \label{uvin}
\begin{aligned}
    v_{\mathrm{in}} &= \int_0^{t_-}\frac{1}{a(\tilde{t}_-)}\dd \tilde{t}+x \,, \\
     u_{\mathrm{in}} &= \int_0^{t_-}\frac{1}{a(\tilde{t}_-)}\dd \tilde{t}-x \,.
\end{aligned}
\end{equation}
Importantly, the coordinates in Eq. \eqref{uvin} satisfy $\dd (\dd v_\mathrm{in})=0$, $\dd (\dd u_\mathrm{in})=0$. With this setup, we are ready to extend the coordinates defined in Eq. \eqref{uvcompactout} into the interior. This is achieved by defining
\begin{equation}
\begin{aligned}
    \bar{v}(t_-,x) &= \bar{v}\!\left[T_+\!\left(T_-\!\left(v_{\mathrm{in}}(t_-,x)\right)\right)\right], \\
    \bar{u}(t_-,x) &= \bar{u}\!\left[T_+\!\left(T_-\!\left(u_{\mathrm{in}}(t_-,x)\right)\right)\right].
\end{aligned}
\end{equation}
Here $\bar{v}(T_+)$ denotes the $\bar{v}$-coordinate trajectory of the thin shell, while $T_-(v_{\mathrm{in}})$ is the inverse of the $v_\mathrm{in}$-coordinate trajectory of the thin shell {$v_{\mathrm{in}}(T_-,x)\big|_{x=x_\mathrm{shell}}$, where $x_\mathrm{shell}=R(T_-)/a(T_-)$.} Completely analogous definitions apply to $\bar{u}(t_-,x)$. See \cite{Bobula:2024ywp} for a general discussion of this type of approach.

The numerically computed conformal diagram covering both the interior FLRW region and the exterior effective Schwarzschild geometry, with the non-isolated thin shell constituting their boundary, is shown in the right panel of Fig. \eqref{conformal}. Note that the causal structure of the effective Schwarzchild vacuum, in context of Oppenheimer-Snyder collapse, was already studied in detail in \cite{Bobula:2023kbo, Lewandowski:2022zce, Fazzini:2023scu}.

\subsection{Comparison with other existing models}

In this subsection, we compare the results of the framework introduced in this work with other extensions discussed in the literature. Our goal is to highlight the advantages—and the inherent complexities—of our approach in extending spacetime beyond shell-crossing singularities. As explicitly demonstrated in the previous section, the non-isolated thin shell formed post-SCS possesses a continuous induced metric and exhibits timelike behavior throughout its entire evolution, while exchanging mass with the expanding FLRW interior.

This represents a fundamental departure from the models presented in \cite{Husain2022,Fazzini:2025hsf,Liu:2025fil}. In those works, either the induced metric is discontinuous at the shock \cite{Husain2022,Fazzini:2025hsf}—displaying two distinct signatures as measured from either side during certain dynamical phases—or it remains continuous but undergoes a signature change during the evolution \cite{Liu:2025fil}. Therefore, a consistent timelike signature cannot be maintained at all times, even though it is a necessary condition for a physically sound thin-shell dynamics.

The origin of this limitation lies in the implicit requirement of Painlevé-Gullstrand time continuity within the integral approach, adopted in the aforementioned works. This requirement can be viewed as an alternative, and mutually exclusive, condition to signature preservation; however, it remains physically unjustified (as shown in \cite{Fazzini2025}). Specifically, if the metric is allowed to be discontinuous across the shock, consistency suggests that the flow of time should also be allowed to differ between the two sides. Even when the induced metric is assumed to be continuous \cite{Liu:2025fil}, the physical distinction between the interior and exterior regions (e.g., a Minkowski interior vs. a Schwarzschild exterior for an isolated shell) implies that the respective time flows should be different. Consequently, PG time slices ought to be allowed to develop discontinuities at the shock—a feature implicitly forbidden by the integral approach, which enforces a continuous time coordinate globally. This argument is further supported by the fact that forcing PG time continuity results in a thin shell that changes its causal character (transitioning from timelike to spacelike via a null stage), which is unphysical for a shell composed of dust particles.

As a result, the global dynamics of the thin shell ({shock} of the gravitational field) differs significantly across these models: 
\begin{enumerate}[(\roman*)]    \item In \cite{Husain2022,Fazzini:2025hsf}, the spacelike nature of the shell motion (as measured by the exterior geometry) allows it to emerge from the black hole apparent horizon, thereby affecting the future of the first asymptotic region.

\item In \cite{Liu:2025fil}, the motion is confined within the non-trapped internal region, and the shock never exits the black or white hole horizons.

\item In our framework, because the shell motion is strictly timelike (consistent with the Oppenheimer-Snyder and isolated thin-shell scenarios), the system—comprising both the interior and the thin shell—evolves toward the second asymptotic region, eventually emerging from the white hole horizon. This is the only possible outcome for the expanding post-bounce timelike motion of stellar matter within this model, due to the time-like requirement for matter evolution.
\end{enumerate}
A further distinction concerns the evolution of the interior matter content. While in \cite{Husain2022,Fazzini:2025hsf,Liu:2025fil} the matter core is rapidly absorbed by the dust thin shell, in our scenario, a fraction of the original stellar mass is retained within the interior throughout the evolution, despite the continuous exchange of mass. This discrepancy is attributed to the different relative motions between the interior and the thin shell across the various models.

Finally, the methodologies used to derive these dynamical solutions differ. The integral approach \cite{Husain2022,Fazzini:2025hsf,Liu:2025fil} treats the system as an {IVP}, where the dynamics is uniquely determined once the specific integral form of the equations is adopted. While our framework also allows for an IVP formulation in principle---{see Figs. \ref{schematic_plot} and \ref{conformal}}—with the Israel junction conditions entering the equations only upon the formation of the SCS—the inherent complexity led us to focus on the post-SCS dynamics by assuming the shell's presence at the initial time {$T_\mathrm{ini}$}. This necessitated careful calibration of the initial conditions, which are fixed post-stellar bounce, whereas weak solutions are typically imposed in the classical regime {long before SCS formation}. Although this might appear as a limitation, the current framework {is expected to be} compatible with a full IVP formulation. In this sense, going beyond the toy model considered here—namely, by implementing in full detail the strategy outlined in Fig. \eqref{schematic_plot} for, e.g., a Gaussian initial dust profile—we expect the junction surface (the non-isolated thin shell), to exist uniquely,  with no freedom in choosing initial conditions at the shock origin. A detailed analysis of this more realistic scenario is, however, left for future work.

\section{conclusions}

One of the primary objectives of any candidate theory of quantum gravity is to describe stellar collapse while avoiding the formation of strong singularities. This work contributes to this endeavor by providing a robust framework and an explicit realization of a regular dynamical evolution.

In the context of effective stellar collapse inspired by loop quantum gravity, models typically predict a core bounce at Planckian densities. However, this is often followed by the formation of shell-crossing singularities, which, although weak in the Tipler sense, pose a challenge for the global extension of the manifold. The central problem addressed in this paper is the extension of such spacetimes beyond the SCS in a physically consistent manner. We interpret the shell-crossing singularity as an infinitesimal, non-isolated dust thin layer, evolved via the Darmois-Israel junction conditions within a Hamiltonian formulation. Our main result is the derivation of a timelike dynamics characterized by a continuous induced metric for the thin shell (acting as a gravitational {shock}). Crucially, this holds for any initial stellar mass, offering a physically sound alternative to the integral approaches frequently adopted in the literature.

{ In pursuing our goal, we proposed a general framework—outlined in Fig.~\ref{schematic_plot}—for constructing a consistent dynamical extension beyond SCS. The proposed strategy does not rely on a specific form of the (classical or effective) equations of motion, instead, it provides a precursor to a detailed and fully consistent analysis of the initial-value problem in any gravitational theory characterized by SCS. }

In our toy model, which serves as a simplified realization of the proposed framework, the formation of the SCS is modeled as a thin dust shell of negligible mass, permitted to exchange matter with a homogeneous interior while maintaining a vacuum exterior. While this setup represents an approximation of the full dynamics—which should ideally emerge from an initial value problem where junction conditions are triggered only upon the first SCS formation—it serves as a powerful heuristic to capture the essential features of the post-SCS evolution. We observe that the shell rapidly accumulates a significant fraction of the interior mass due to the non-zero relative velocity between the shell and the dust fluid. Consequently, the system expands toward the anti-trapped region of the maximally extended effective vacuum, eventually emerging from a white hole horizon into a secondary asymptotic region. This evolution describes an inter-universal transition, wherein the star traverses the throat with Planckian curvature of a wormhole-like structure.

Despite resolving the limitations of the integral method, the emerging picture introduces further theoretical challenges. Specifically:

\begin{enumerate}[(\roman*)]     \item \textit{Mass Inflation}: The existence of a Cauchy horizon, analogous to the Reissner-Nordstr\"om solution, suggests a susceptibility to mass inflation instability. As investigated in \cite{Cao:2023aco}, this effect may lead to a null physical singularity on the inner horizon. 

\item \textit{Loss of predictability}: The presence of a Cauchy horizon, as exemplified in the Oppenheimer-Snyder scenario {\cite{Bobula:2023kbo, Lewandowski:2022zce, Fazzini:2023scu}}, generally entails a loss of global predictability. While a model transcending the current effective framework might ultimately resolve this issue, its implications remain a fundamental challenge that cannot be overlooked.

\item \textit{Eardley Instability}: The classical Eardley instability \cite{Eardley:1974zz} is expected to affect the effective white hole horizon. For macroscopic stars, this could trigger a recollapse immediately upon emergence, potentially masking any astrophysical signatures in the second asymptotic region.

\item \textit{Information paradox}: The present analysis neglects Hawking radiation. In the effective framework, the collapse leads to an eternal black hole from the perspective of an observer living in the same asymptotic region of the collapsed star. However, accounting for evaporation
implies that the black hole mass should eventually dissipate, leading potentially to the information paradox, similarly to the classical case.

\end{enumerate}

In principle, the aforementioned issues could be circumvented by incorporating Hawking radiation into the dynamics, while the information paradox might be addressed through arguments such as those presented in \cite{Rovelli:2017mzl}. In the underlying effective model, the trapped region shrinks and eventually vanishes as the black hole mass approaches the Planck scale \cite{Kelly:2020uwj}. This scenario offers a compelling alternative to the black-to-white hole transition paradigm \cite{Rovelli:2024sjl}, which may prove unnecessary for this specific model. Indeed, since the trapped region disappears naturally at the end of the evaporation process, there is no need for ad hoc metric gluing \cite{Han:2023wxg}—a construction that remains external to the solutions of the effective equations.

While including Hawking evaporation is beyond the scopes of this work and will be the aim of future research, the framework developed herein provides a consistent and important stepping stone for investigating the fate of stellar matter beyond shell-crossing singularities, laying the groundwork for more comprehensive semi-classical models.

\section*{Acknowledgements}
MB acknowledges support from the Polish National Science Centre (Narodowe Centrum Nauki, NCN) under grant OPUS 2020/37/B/ST2/03604.

\appendix

\section{Gauge fixing and polymerization of constraints}

In this appendix, we derive the polymerized Hamiltonian constraint and integrate it across the thin shell. We begin by expressing the LTB metric, for both the interior and exterior regions of the shell, in terms of the Ashtekar-Barbero variables
\begin{equation}
\dd s^2= -N_\pm^2\dd t_\pm^2 +\frac{(E^b_\pm)^2}{E_\pm^a}(\dd r_\pm+ N^r_\pm \dd t_\pm)^2 + E_\pm^a \dd \Omega_\pm^2    ~,
\end{equation}
where $E^a$ and $E^b$ correspond to the radial and angular components of the densitized triad, respectively \cite{Husain2022}, while $N$ and $N^r$ denote the lapse function and the radial component of the shift vector. We assume that the coordinate $r$ (which, upon imposing the areal gauge, satisfies $r=\sqrt{E^a_\pm}$) is continuous across the shell, in accordance with the first Darmois-Israel junction condition. Consequently, we adopt a unified notation $r$ for both $r_+$ and $r_-$.

The core principle of the canonical formulation for the second Darmois-Israel junction condition is to solve the integrated version of the constraints \eqref{diffeo}, allowing for gauge fixing prior to integration \cite{Crisostomo:2003bb}. In the following, the + $(-)$ superscripts or subscripts denote the regions $r\geq R$ ($r < R$), respectively, with $R$ denoting the shell location at generic time.

We begin by considering the classical setup, where the constraints in \eqref{diffeo} take the following form:
\begin{align}
& \mathcal{H}_t^\mathrm{grav.}={^\pm \mathcal{H}}^\mathrm{grav.}_t ~,\\
 &  \mathcal{H}_r^\mathrm{grav.}={^\pm\mathcal{H}}^\mathrm{grav.}_r,  \\
 & \mathcal{H}^\mathrm{matt.}_t=4 \pi \sqrt{p_T^2+\frac{E^a_-}{(E^b_-)^2}p_T^2(\partial_r T)^2  }\cdot \theta(R-r)~   , \\
 &\mathcal{H}^\mathrm{matt.}_r=-4\pi p_T \partial_r T \cdot \theta(R-r) ~,
\end{align}
with
\begin{align}
{^\pm\mathcal{H}}^\mathrm{grav.}_t   =&-\frac{1}{2G\gamma}\left[\frac{2ab\sqrt{E^a}}{\gamma}+\frac{E^b}{\gamma \sqrt{E^a}}(b^2+\gamma^2)\right]_\pm- \notag\\
&-\frac{1}{2G}\left[\frac{ (\partial_r E^a)^2}{4E^b \sqrt{E^a}}+\sqrt{E^a}\partial_r\left( \frac{\partial_r E^a}{E^b}\right)  \right]_{\pm} ,  \end{align}
\begin{align}
{^\pm \mathcal{H}}^\mathrm{grav.}_r= \frac{1}{2G\gamma}[2E^b \partial_r b-a\partial_r E^a]_\pm ~,
\end{align}
where here $a(r,t)$ and $b(r,t)$ represent the radial and angular components of the extrinsic curvature \cite{Husain2022} (which, together with the spin connection, forms the Ashtekar-Barbero connection). The explicit contribution of the thin shell to the constraints is omitted here, as it is not strictly required for the subsequent derivation.

The variable $T$ denotes the dust field in the interior, with $p_T$ as its conjugate momentum; the Heaviside functions ensure that the dust contribution is confined to the interior spacetime $(r<R)$, consistent with the assumption of a vacuum exterior. Rather than integrating the constraints and requiring the result to vanish weakly, we simplify the procedure by first imposing the areal gauge and the diffeomorphism constraint. We then solve the integrated version of the resulting Hamiltonian constraint, following the approach in \cite{Crisostomo:2003bb}. Furthermore, while the system can be gauge-fixed using the dust-time gauge prior to integration, the resulting Hamiltonian constraint must be solved post-integration.

We proceed by imposing the areal gauge in both the interior and exterior regions, setting $E^a_\pm=r^2$. This choice enables us to enforce the non-integrated version of the diffeomorphism constraint, which effectively eliminates the variable $a$ from the Hamiltonian constraint
\begin{align}
&\mathcal{H}^\mathrm{grav.}_t= 
N^r_\pm (\mathcal{H}^\mathrm{shell}_r+\mathcal{H}^\mathrm{matt.}_r)- \notag\\
&-\frac{1}{2G\gamma}\left[\frac{E^b (\partial_r b)^2}{\gamma}+ \frac{E^b}{\gamma r}(b^2+\gamma^2)-\frac{\gamma r}{E^b}-\gamma r \partial_r \left( \frac{2r}{E^b}\right) \right]_\pm,
\end{align}
where $N^r_\pm=-\frac{b_\pm}{\gamma}$ is obtained by requiring that the areal gauge be dynamically preserved in the interior and exterior regions, respectively \cite{Husain2022}. Furthermore, we set $E^b_\pm=r$ (corresponding to the solution of the equations of motion for both regions \cite{Husain2022}) to achieve the desired marginally bound Painlevé-Gullstrand form, as expressed in \eqref{FLRW_PG} and \eqref{pg+}. The full Hamiltonian constraint then takes the form:
\begin{align}
\mathcal{H}_t=& N^r_\pm(\mathcal{H}^\mathrm{shell}_r+\mathcal{H}^\mathrm{matt.}_r)+\mathcal{H}^\mathrm{matt.}_t+\mathcal{H}^\mathrm{shell}_t ~-\notag \\
&-\frac{1}{2G \gamma^2}\left[ \partial_r (rb^2)\right]_\pm.
\end{align}
At this stage, we impose the dust time gauge $T_- -t_-=0$ for the interior, whose dynamical consistency fixes $N_-=1$ \cite{Husain2022}, while setting $N_+=1$ for the exterior \cite{Kelly:2020uwj}. In the interior, the dust-time gauge implies that the matter contributions to the constraints are $\mathcal{H}^\mathrm{matt.}_t=4\pi p_T$, and $\mathcal{H}^\mathrm{matt.}_r=0$. Finally, we recall that the dust contribution to the scalar constraint, integrated over the angular directions, can be recast in terms of the dust energy density
\begin{equation}
    \mathcal{H}^\mathrm{matt.}_t=\int \dd \theta \int \dd \phi \rho \sqrt{q_-}=4 \pi \rho \sqrt{E^a_-} E^b_-=4 \pi \rho r^2 ~,
\end{equation}
where $q_-$ denotes the determinant of the spatial metric in the $(-)$ region, and the Heaviside function has been omitted for brevity. Here, we have defined $\rho$ as the dust energy density of the stellar interior; this quantity should be clearly distinguished from the surface density of the thin shell. We thus obtain
\begin{align}
 &\mathcal{H}_t=-\frac{\left[ \partial_r (rb^2)\right]_\pm}{2G \gamma^2}+N^r_\pm \mathcal{H}^\mathrm{shell}_r+ 4\pi\rho r^2\theta(R-r)+\mathcal{H}^\mathrm{shell}_t ~.
\end{align}
Notice that, although the time gauge has been fixed for both the interior and exterior regions, we cannot solve the Hamiltonian constraint locally before performing the integration. Doing so would fail to properly enforce the integrated version of the constraint across the shell. At this stage, we implement the polymerization of the geometric sector of the fully gauge-fixed Hamiltonian constraint and the shift vector, following the effective $\bar{\mu}+K$-loop quantization prescription \cite{Husain2022}:
\begin{align}
 & b_\pm \rightarrow \frac{r}{\sqrt{\Delta}}\sin\frac{\sqrt{\Delta}b_\pm}{r},\\
& N^r_\pm \rightarrow -\frac{r}{2 \gamma \sqrt{\Delta}}\sin\left( \frac{2 \sqrt{\Delta}b_\pm}{r}\right) ~,
\end{align}
and subsequently integrating across the shell surface, to obtain
\begin{align}
&\int_{R-\varepsilon}^{R+\varepsilon}\mathcal{H}^\mathrm{eff}_t \dd r=-\frac{1}{2G \gamma^2} \frac{r^3}{\Delta}\sin\left( \frac{\sqrt{\Delta}b}{r}\right)^2\bigg|_{R-\varepsilon}^{R+\varepsilon}+\\
&+\int_{R-\varepsilon}^{R+\varepsilon}\dd r (N^r_{\pm} \mathcal{H}^\mathrm{shell}_r+\mathcal{H}^\mathrm{shell}_t )\approx 0  ~, 
\end{align}
where the term proportional to the dust energy density $\rho$ vanishes upon integration between $R-\varepsilon$ and $R$ in the limit $\varepsilon \rightarrow 0$. By substituting the effective LTB solutions for $b$ in the FLRW interior and the Schwarzschild exterior \cite{Husain2022}, we finally arrive at:
\begin{equation}
M-\frac{4}{3}\pi \rho R^3= - \lim_{\varepsilon \rightarrow 0}\int_{R-\varepsilon}^{R+\varepsilon} \dd r (N^r_{\pm}\mathcal{H}^\mathrm{shell}_r+\mathcal{H}^\mathrm{shell}_t )~,
\end{equation}
where on the left-hand side we performed the $\varepsilon \rightarrow 0$ limit.

\label{appendix0}
\section{covariant conservation of $T^{\mu \nu}$ across the shell}

\label{appendix1}

In this section, we demonstrate that covariant conservation across the shell leads to the expression \eqref{consmass}. We begin by considering the energy-momentum tensor for a spacetime containing a thin layer (separating a FLRW interior from an effective vacuum exterior). In Gaussian normal coordinates $\{ \tau, \chi, \theta, \phi \}\equiv \{ \chi, x^i\}$, where $\chi$ denotes the spatial coordinate normal to the shell, the metric takes the general form:
\begin{equation}
 \dd s^2= h_{ij}(\chi, x^i) \dd x^i \dd x^j + \dd \chi^2   ~,
\end{equation}
and $h_{ij}(0,x^i)={^\pm e}^\mu_i {^\pm e}^\nu_j g^\pm_{\mu \nu}$ is the induced metric on the shell hypersurface. As in the classical formulation \cite{Poisson2004Toolkit}, the explicit computation of $h_{ij}$ is not required. While a continuous map across the hypersurface is necessary to establish the formal framework, the final result will be expressed in terms of the charts $\left\{x^\mu_\pm \right\}$.

The energy-momentum tensor for the spacetime with a thin shell at $\chi=0$ reads
\begin{equation}
T^{\mu \nu}= T^{\mu \nu}_+ \theta(\chi)+T^{\mu \nu}_- \theta(-\chi)+S^{\mu \nu}\delta( \chi) ~, 
\end{equation}
where the coordinates are continuous across the shell, but cover only a local neighborhood of the hypersurface. 

Furthermore, $T^{\mu \nu}_-=\rho u_F^\mu u_F^\nu$, where $\rho$ is the proper energy density of the interior fluid, and $u_F^\mu$ is the FLRW fluid $4$-velocity. In the exterior region, we have $T^{\mu \nu}_+=0$ as it is assumed to be vacuum. The surface energy-momentum tensor is given by $S^{\mu \nu}=\sigma u^\mu u^\nu$, where $\sigma$ denotes the proper surface energy density of the shell and $u^\mu$ is the 4-velocity of the shell in gaussian normal coordinates. It is important to note that, in general, the components of the same 4-vector will differ between charts $\chi^\mu \neq \chi^\mu_\pm$, and similarly and $u^\mu \neq u^\mu_\pm$. In particular, the normal vector and the 4-velocities, as measured in the interior and exterior frames, do not match continuously across the shell hypersurface.

We now impose the covariant conservation of the total energy-momentum tensor
\begin{equation}
 \nabla^\mu T_{\mu \nu}= T^-_{\mu \nu} \nabla^\mu \theta(-\chi)+\nabla^\mu(S^-_{\mu \nu}\delta(\chi) )  =0 ~,
    \label{conservative}
\end{equation}
where we have omitted the term $\nabla^\mu T^-_{\mu \nu}$, as it represents the energy-momentum conservation in the interior bulk, which is identically satisfied. To proceed, we contract this expression with the shell $4$-velocity $u^\mu$ in gaussian normal coordinates, noting that
\begin{align}
 T_{\mu \nu}^- u^\nu&= (u_F \cdot u) \rho u_{\mu}^F=(u_F\cdot u) \rho g_{\mu \nu }u^\nu_{F}=\notag \\
& =(u_F \cdot u) \rho (h_{\mu \nu} - \chi_\mu \chi_\nu-u_\mu u_\nu) u_F^\nu=   \notag \\
&=(u_F \cdot u) \rho [-\chi_\mu (\chi \cdot u_F) - u_\mu(u\cdot u_F) ] ~,
\end{align}
where in the last step we used the fact that $u_F$ possesses no angular components, while $h_{\mu \nu}$ acts as the projector onto the angular subspace. Substituting this into \eqref{conservative}, we obtain
\begin{equation}
 \delta(\chi) \rho (u_F \cdot u) ( u_F \cdot \chi) +u^\nu \nabla^\mu (S_{\mu \nu} \delta(\chi))=0   ~,
\end{equation}
where we used the identities $\dd \theta(\chi) / \dd\tau =0$, and $\dd\theta(-\chi) /\dd \chi =-\delta(\chi) $. Applying the Leibnitz rule and noting that $\dd \delta(\chi) / \dd\tau =0$ (since the shell position is fixed at $\chi=0$ in these coordinates), we obtain
\begin{equation}
  u^\nu (\nabla^\mu S_{\mu \nu})\delta(\chi)=-\delta(\chi) \rho (u_F \cdot u) (u_F \cdot \chi) ~.
\end{equation}
Recalling that $u^\mu={e^\mu_\tau}$, we have
\begin{equation}
  {e^\nu_\tau} \nabla^\mu S_{\mu \nu} =- \rho (u_F \cdot u) (u_F \cdot \chi) ~.
  \label{cons2}
\end{equation}
By using $\nabla_\mu e^\nu_a =0$, we get
\begin{equation}
 e^\nu_ a \nabla_\mu S^\mu_{\nu}=D_b S^b_a ~, 
 \label{useful}
\end{equation}
where $S^{\mu \nu}=S^{ab}e^\mu_a e^\nu_b$ and $D_a$ is the intrinsic derivative on the shell hypersurface. Therefore
\begin{equation}
 D_\tau S^\tau_\tau=-\rho (u_F \cdot u) (u_F \cdot \chi)=-\rho (u_F \cdot u^-) (u_F \cdot \chi^-)~,
\end{equation}
 since $S$ has only $\tau\tau$ component: $S^a_b=\mathrm{diag}(\sigma, 0,0)$. In the last line we used the invariance of the scalar product under change of coordinates, and we have written the two quantities in the coordinate system $\{t_-,r, \theta, \phi  \}$ of the interior.
 
 After a short computation, one then can evaluate the covariant derivative of $S$ on the hypersurface, getting
\begin{align}
 \frac{\dd (4 \pi R^2 \sigma) }{\dd \tau}&=-4\pi R^2 \rho (u_F \cdot u)^- (u_F \cdot \chi)^-= \notag\\
 &=- R^2 \rho(T_-) \dot{T}_- (\dot{R} +\dot{T}_- N^r_-  ) ~,
\end{align}
if the interior is written in Painlevé-Gullstrand coordinates.

\label{appendix1}

\end{document}